\documentclass[traditabstract]{aa} % for the abstract without structuration 
\usepackage{natbib}

\usepackage{graphicx}
\usepackage{txfonts}
\begin{document}

   \title{Very Large Telescope observations of {\it Gomez's Hamburger:}  \\ Insights into a  young protoplanet candidate \thanks{Based on observations collected at the European Southern Observatory, Chile under program ID 385.C-0762A.}
}

   \author{O. Bern\'e \inst{1,2}
        \and 
        A. Fuente\inst{3}
        \and
        E. Pantin\inst{4} 
        \and
        V. Bujarrabal\inst{3}
        \and 
        C. Baruteau\inst{1,2}
        \and
        P. Pilleri\inst{1,2}
          \and
          E. Habart\inst{5}
          \and
          F. M\'enard\inst{6,7,8}
          \and
        J. Cernicharo\inst{9,10} 
          \and
         A. G. G. M. Tielens\inst{11}
          \and
        C. Joblin\inst{1,2}
          }
          
\titlerunning{VLT observations of GoHam}

   \institute{ 
          Universit\'e de Toulouse; UPS-OMP; IRAP;  Toulouse, France
                 \and CNRS; IRAP; 9 Av. colonel Roche, BP 44346, F-31028 Toulouse cedex 4, France
         \and Observatorio Astron\'omico Nacional, Apdo. 112, 28803 Alcal\'a de Henares, Madrid, Spain.
         \and Service d'Astrophysique CEA Saclay, France
         \and Institut d'Astrophysique Spatiale, Paris-Sud 11, 91405 Orsay, France
         \and Millenium Nucleus ``Protoplanetary Disks in ALMA Early Science,'' Universidad de Chile, Casilla 36-D, Santiago, Chile
         \and UMI-FCA 3386, CNRS/INSU, Casilla 36-D, Santiago, Chile
         \and Univ. Grenoble Alpes, IPAG, F-38000 Grenoble, France\\
  CNRS, IPAG, F-38000 Grenoble, France
         \and Instituto de Ciencia de Materiales de Madrid (ICMM-CSIC). Sor Juana Ines de la Cruz 3, 28049 Cantoblanco, Madrid, Spain.
         \and Centro de Astrobiolog\'{\i}a, CSIC-INTA, Ctra. de Torrej\'on a Ajalvir km 4, E-28850 Madrid, Spain.
          \and Leiden Observatory, Leiden University, Niels Bohrweg 2, NL-2333 CA Leiden, The Netherlands. 
         }

   \date{}

% \abstract{}{}{}{}{} 
% 5 {} token are mandatory
 
  \abstract
  % context heading (optional)
  % {} leave it empty if necessary  
   {Planets are thought to form in the gas and dust disks around young stars. In particular, it has been proposed that 
   giant planets can form through the gravitational instability of massive extended disks around intermediate-mass stars. However,
   we still lack direct observations to constrain this mechanism. 
   We have spatially resolved the 8.6 and 11.2 $\mu$m  emission of a massive protoplanetary disk seen edge on around an A star,  
   \emph{Gomez's Hamburger} (GoHam), using VISIR at the Very Large Telescope. A compact region situated 
    at a projected distance of $350\pm50$ AU south of the central star is found to have a reduced emission.
  This asymmetry is fully consistent with the presence of a cold density structure, or clump, identified 
    in earlier CO observations, and we derive physical characteristics consistent with those observations:  a mass of
    0.8-11.4 Jupiter masses (for a dust-to-gas mass ratio of 0.01), a radius of about 10$^2$ astronomical units, and a local density 
    of about $10^{7}$ cm$^{-3}$. 
    %Observationally, GoHam b is comparable to pre-stellar cores found embedded in molecular clouds, but in a smaller and denser version, embedded in a molecular disk.  
    Based on this evidence, we argue that this clump, which we call GoHam b, is a promising candidate for a young protoplanet formed by gravitational instability that might be representative of the precursors of massive planets observed around A stars, such as HR 8799 or Beta pictoris. More detailed studies at high angular resolution are needed  to better constrain the physical properties of this object to confirm this proposal.}

   \maketitle
%
%________________________________________________________________

\begin{figure*}[!ht]
\begin{center}
\includegraphics[width=18cm]{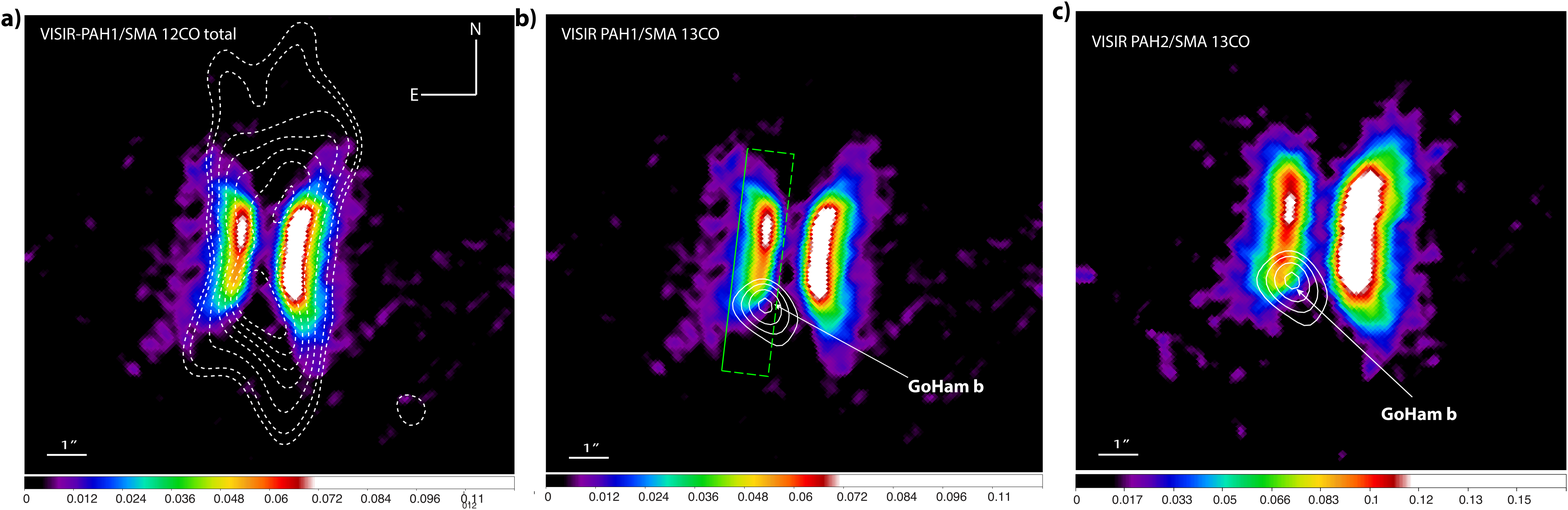}

\vspace{-0.0cm}
\caption{{ {\bf a)} VLT-VISIR 8.6 $\mu$m (PAH1 filter) image of GoHam in color, the scale is in Jy/arcsec$^2$. In contours : velocity-integrated $^{12}$CO(2-1) emission observed with the SMA from \citet{buj08, buj09}. {\bf b)} VLT-VISIR 8.6 $\mu$m (PAH1 filter, same as left panel) in color. Contours show the emission of $^{13}$CO (2-1) emanating from from GoHam b after subtracting the best-fit disk model (see \citealt{buj09} for details). This region also corresponds to the local decrease of mid-IR emission seen in the VISIR image. The position of the cross cut used to extract the profiles shown in Fig.~\ref{fig_cut} is shown in green.
{\bf c)}  VLT-VISIR 11.2 $\mu$m (PAH2 filter) image of GoHam in color, the scale is in Jy/arcsec$^2$. Contours as in b).
%{\it Hubble}-NICMOS image of GoHam, intensity scale is in DN units (multiplied by 2.86e-19 gives ergs/cm$^{-2}$/$\AA$). Contours as in {\bf b)}.
}
 \label{fig_visir}}
\end{center}
\vspace{-0.5cm}
\end{figure*}

\section{Introduction}

The disks present around low- to intermediate-mass young stellar objects have gained much interest
in recent years since they are believed to be the cradles of planetary formation. 
In this context, the study of their dust and gas content is crucial because dust is the primary reservoir 
of matter available to form telluric planets and the cores of giant planets.
The recent discoveries by direct-imaging techniques of over 20 giant planets with orbital separations 
from 10 to a few hundred AU suggest that giant planets can form in the outer regions of protoplanetary
disks ({ e.g.,  Beta pictoris and HR 8799 with planets at estimated separations between $\sim $9 and 
70 AU, respectively; \citealt{lag10, mar10}}). In these regions disks can become unstable to their own gravity and 
form clumps of molecular gas that can continue to accrete mass and later evolve to become planets 
(or brown dwarfs). This gravitational instability (GI, e.g., \citealt{bos97}) scenario is a possible channel 
for planet formation around A or B stars, but direct observational evidence is scarce. 
One possible example of a candidate protoplanet that might have
been formed by GI was recently provided by \citet{qua13, qua14}, who identified a source in the disk around the Herbig star HD 100456
at a projected separation of $\sim$ 50 AU.  The mass of this source could not be derived directly,  but 
is at most a few Jupiter masses ( \citealt{boc13, cur14}).

Gomez's Hamburger (IRAS 18059-3211; hereafter GoHam) is an A star surrounded by  
a dusty disk. When first studied by \citet{rui87}, it was classified as an evolved object (post-AGB star) on the basis 
of its spectral type and the presence of dust. However, all recent studies \citep{buj08, woo08, buj09, deb10} 
clearly indicate that it is a young A star surrounded by a protoplanetary disk. The distance to GoHam 
is not known precisely, but a value $d=250 \pm50$ pc is required to satisfy all the existing observational
constraints \citep{woo08, ber09, buj09}. We here adopt this value with the uncertainty.
GoHam presents intense CO emission; SMA maps of $^{12}$CO and $^{13}$CO J=2-1 lines very clearly show the Keplerian dynamics 
of the disk \citep{buj08, buj09}. The lower limit for the disk mass derived from these CO observations
is of about 10$^{-2}$ M$_{\sun}$, while the mass upper limit is estimated to be $\sim$ 0.3 M$_{\sun}$ 
based on dust emission \citep{buj08, woo08} and assuming an interstellar dust-to-gas mass ratio of 0.01.
Overall, GoHam appears to be similar to isolated Herbig stars \citep{mee01}
such as HD 100546 and HD 169142, in a more massive version, but still smaller 
than the recently discovered disk around CAHA J23056+6016 \citep{qua10}.
%These are often considered as precursors of massive 
%planetary systems around intermediate mass stars such as Beta Pictoris,  HR 8799 or HD 95086. 
GoHam is seen almost perfectly edge on, which offers the possibility to study this class of objects from a new
and complementary perspective, in particular, with improved constraints on the vertical structure
of the disk. Using a radiative transfer model to predict line emission from a Keplerian flaring disk, \citet{buj09}
derived a large-scale description of the physical conditions throughout the disk. After subtracting the 
model that best fit the observations,  these authors found a significant  residual emission situated about 1.3'' ($330\pm70$ AU) 
south of the central star, which they identified as a gas condensation, containing a mass between one and few times that 
of Jupiter. Hence, this source was proposed to be a candidate 
protoplanet, possibly resulting from a GI collpapse. 

GoHam shows bright polycyclic aromatic (PAH) emission \citep{woo08, ber09}, as often observed toward
Herbie Ae/Be stars \citep{ack04, hab04}. In this letter, we present high angular resolution imaging
of GoHam obtained in the PAH filters with VISIR at the Very Large Telescope, which provides a new view of
the vertical structure of the disk.

\section{Observational results}

\subsection{Observations}

GoHam was observed  with VISIR at the Very Large Telescope. We obtained three
exposures in chopping-nodding in the PAH1 filter (8.6 $\mu$m) on April 22-23, 2010,
and two exposures in the PAH2 filter (11.3 $\mu$m) on August 29 and September 01, 2010. 
The total integration time was 1552 seconds for the PAH1 filter and 1702 seconds for the 
PAH2 filter. Observations in the PAH1 filter were conducted with an exceptional seeing 
of 0.5'' , while the observations in the PAH2 filter were conducted with a seeing 
ranging between 1.0 and 1.6''. Calibration was achieved using the ESO-provided observation 
of the standard star HD 177716. 
%The 10$\sigma$ one hour sensitivity per pixel is 4.43 mJy. 
In addition, we used the submillimeter array (SMA) data for the $^{12}$CO(2-1) and 
$^{13}$CO(2-1) lines observed in June 2006 that were presented in \citet{buj09} with 
a beam size of $\sim 1.1\times1.5''$.
%We have also retrieved from the online archive the near infrared 1.1 $\mu$m  image of GoHam observed 
%with NICMOS (FW110 filter) onboard the {\it Hubble} space telescope.
The VISIR and part of the SMA  observations are shown in Fig. \ref{fig_visir}.

%We have observed GoHam with SPIRE onboard \emph{Herschel} in the single pointing,
%high spectral resolution mode with a repetition factor of 60. We ran the pipeline using HIPE to obtain Level 2 data
%and the final apodized spectrum is shown in Fig. \ref{fig1}. The sensitivity is of the order of a few $10^{-18}$
%W.m$^{-2}$.

%\section{Results}

\subsection{Disk morphology as seen with PAHs}

The images obtained in the PAH1 and PAH2 filters are shown in Fig. \ref{fig_visir}. 
The observed mid-IR emission results from UV-excited polycylic aromatic hydrocarbons 
(PAHs, \citealt{tie08}). The PAH1 filter covers part of the C-C vibration at 7.7 $\mu$m and the C-H vibration at 8.6 $\mu$m, while
the PAH2 filter mainly covers the C-H vibration at 11.3 $\mu$m (see Fig.~\ref{fig_spitzer}).
The edge-on disk is clearly resolved, and both faces
are separated by a dark lane, where the disk becomes optically thick to its own mid-IR light.
This lane is 1.5'' broad, that is, $375\pm75$ AU, larger than what can be observed for any other disk 
in the mid-IR and indicating the massive nature of GoHam.
Both images show an asymmetry in the fluxes found on the western
and eastern faces of the disk caused by the slight inclination of the disk to the line of sight ($\sim 5-10^{\circ}$).
The radial extent of the PAH emission ($R_{PAH} \sim$3'' , i.e., $750\pm150$ AU) is much smaller than the 
radial extent observed for the molecular gas in CO (2-1) emission ($R_{mol} \sim$6.6'' , i.e., $1650\pm350$ AU). 
On the other hand, the PAH emission extends to  higher altitudes ($H_{PAH}\sim$ $770 \pm150$ AU) above the disk than 
the CO emission  ($H_{mol} \sim$ $450\pm90$ AU, see \citealt{buj09}).
%This can be explained by the fact that PAHs are emitting in a much warmer gas. 
In photodissociation regions, PAH molecules mainly emit in the warm atomic gas,
as evidenced, for instance, by the strong spatial correlation with the [CII] fine-structure line \citep{job10}. 
Therefore, and since they are dynamically coupled to the gas, PAHs can be
 considered as a tracer of the warm (a few 100 K, \citealt{job10}) external layers 
 of protoplanetary disks. This warm gas is therefore expected to have a higher scale height than the cold molecular gas traced 
 by low-$J$ CO emission. The proposed morphology is shown in  Fig.~\ref{fig_sketch}.

\subsection{Recovery of the GoHam b condensation in absorption in the mid-IR}\label{sec_proto}

\begin{figure}
\begin{center}
\includegraphics[width=9cm]{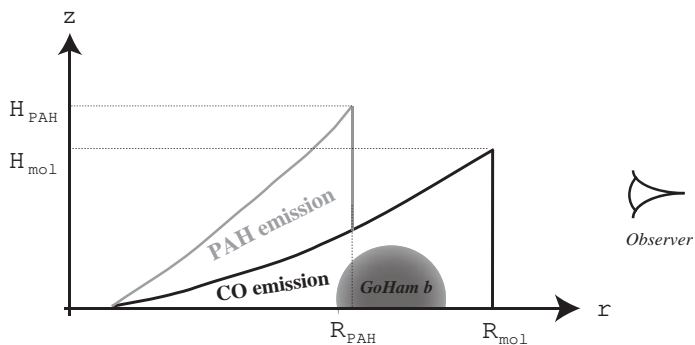}
\vspace{-0.0cm}
\caption{ Schematic representation of the different components of GoHam's disk with the main geometrical parameters. \label{fig_sketch}}
\end{center}
\vspace{-0.5cm}
\end{figure}

Since the PAH emission arises from the surface layers of the flared disk, this PAH emission 
has to traverse the disk before reaching us (Fig.~\ref{fig_sketch}). At high altitudes the crossed disk slice is thin, 
and PAH emission is marginally absorbed by the disk. On the other hand, at lower altitudes the 
PAH emission has to traverse part of the disk and is largely absorbed. 
Absorption at mid-infrared wavelengths is caused by the strong absorption band at 9.7 $\mu$m 
that is a result of the silicates in dust grains. The beam-averaged mid-infrared spectrum of GoHam 
obtained with Spitzer is shown in Fig.~\ref{fig_spitzer}. We fit this spectrum using the 
PAHTAT toolbox\footnote{http://userpages.irap.omp.eu/~cjoblin/PAHTAT/Site/PAHTAT.html}  
, which allows adjusting an observed spectrum using a set of PAH template spectra, an underlying
continuum, and a correction for the extinction that is  due to silicates (see \citealt{pil12} for details). 
The results of this procedure are shown in Fig.~\ref{fig_spitzer} and demonstrate that the average 
spectrum of GoHam is indeed affected by absorption of silicates, with an optical depth of about 0.4-0.5 at the wavelengths of the VISIR filters. 

After subtracting their radiative transfer model, \citet{buj09} found a 
residual gas emission in the disk, identified as a condensation of molecular
gas. Panels b) and c) of Fig.~\ref{fig_visir} show this residual $^{13}$CO emission tracing the gas condensation, 
overlaid on the PAH1 and PAH2 images. The condensation 
clearly spatially corresponds to a local decrease of PAH emission, where the disk is typically less 
bright by 30\% than at the symmetric region in the north.
This asymmetry is clearly visible in the profile (Fig.~\ref{fig_cut}) obtained along the cross cut
shown in Fig.~\ref{fig_visir} and seems to match the position of the CO clump.
Overall, we conclude that we probably detected the condensation reported by \citet{buj09} as absorption 
in the PAH1 and PAH2 filters of VISIR\footnote{Note that, although the SMA and VISIR observations are separated 
by several years, the source has not moved. This is consistent with the orbital period, which is longer than 200
years for a clump on an orbit that is at least 350 AU.}. From an observer's point of view, this condensation
is therefore similar to infrared dark clouds seen against bright mid-IR backgrounds, which are believed to be
pre-stellar cores embedded in molecular clouds \citep{rat06}. GoHam b can be seen as such a core,
but embedded in a molecular disk. In the following, we refer to this condensation 
as GoHam b, following the nomenclature adopted by \citet{qua13}.

\section{Physical properties of GoHam b}\label{sec_prop}

%\subsection{Observational properties}

In the VISIR images (Fig.~\ref{fig_visir}), GoHam b corresponds to a region of increased absorption.
This can be understood as resulting from a localized increase of the dust and gas column density at the position 
where \citet{buj09} detected an excess of CO emission. 
To gain more insight into the properties of GoHam b, we modeled the
mid-IR and $^{13}$CO emission profiles shown in Fig.~\ref{fig_cut} with a simple parametric
model (see Appendix~\ref{app_mod}). The model is adjusted so as to simultaneously reproduce the
$^{13}$CO profile and obtain a mid-IR profile corrected for absorption that is symmetric
(see Fig.~\ref{fig_cut}). With this model, we derive a radius of GoHam b of $155\pm31$ AU,
a density $n_b=7.0\pm1.4\times10^{6}$ cm$^{-3}$ , and a mass $M_b=0.95\pm0.19 $ M$_{Jup}$.
If the dust cross-section per H atom is divided by a factor of 10, as can be the case for protoplanetary disks \citep{and09, dal01},
then the density and mass are higher by one order of magnitude
(i.e., $M_b=9.5\pm1.9 $ M$_{Jup}$ , and $n_b=7.0\pm1.4\times10^{7}$ cm$^{-3}$).
The final mass range is therefore $M_b=0.8-11.4$ M$_{Jup}$, which
agrees well with earlier 
estimates by \citet{buj09}, who found masses ranging between  1 and a few Jupiter masses.

%Finally, we derive a peak column density in the PAH1 image (which 
%was obtained with the best seeing and hence smallest beam dilution) of the order of
%$10\times10^{22}$ cm$^{-2}$ for GoHam b. 

%It would not be surprising that the spectral energy distribution of this type of object also shares similarities with prestellar cores, as is already suggested by the detection of the source in absorption in the mid-infrared.

\begin{figure}
\begin{center}
\includegraphics[width=8cm]{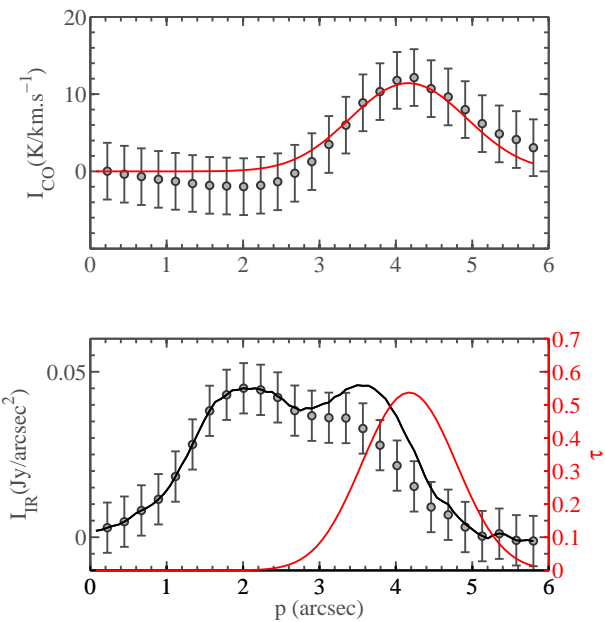}
\label{fig_cut}
\vspace{-0.0cm}
\caption{Emission profiles obtained along the cut shown in Fig.~\ref{fig_visir}. The upper
panel presents the observed $^{13}$CO (2-1) emission of GoHam b (circles with error bars) and the
fit obtained with the model ($I_{CO}$ see Appendix~\ref{app_mod}) in red. The lower panel shows 
the observed mid-IR emission profile ($I_{IR}$) for the same cut (circles with error bars) and the 
mid-IR profile obtained after correction for extinction due to GoHam b ($I_{IR}^0$) in black. The red
curve shows the optical depth  profile $\tau_b$ used for this correction.}
\end{center}
\vspace{-0.5cm}
\end{figure}

\section{GoHam b: a candidate protoplanet ?}

\subsection{Comparison to other substructures seen in disk}

In the recent years, high angular resolution observations have revealed asymmetric features
inside protoplanetary disks. In this context, it is interesting to compare the condensation
seen in GoHam to other structures observed in disks. {  Spiral arms have been observed in several disks
(e.g., \citealt{mut12, cas12, ram12, tan12, gra13, boc13, ave14}), but it seems difficult to understand GoHam b 
as a spiral arm since it is a unique, unresolved, dense, and isolated source in the disk, while spiral arms are generally 
extended and/or multiple and do not necessarily emanate from dense regions.} GoHam b could be an arc or horseshoe structure, as have been 
observed in Oph-IRS 48 \citep{vdm13, bru14}, HD 142527 \citep{cas13}, HD 135344B,
and SR 21 \citep{per14}, for instance. However, this type of asymmetries mostly concerns the emission of dust, whereas here, 
both dust (as traced  by the 8.6 $\mu$m optical depth or the millimeter emission, see Fig. 1 in \citealt{buj08}) 
and gas column densities are higher toward GoHam b. While we cannot rule out hypothetical horseshoe made of
both gas and dust, there is to our knowledge no clear observational evidence for this phenomenon so far.

An asymmetry in the mid-IR PAH emission of the edge-on disk around the Herbig Star PDS 144N 
was reported by \citet{per06} and appears to be very similar to the asymmetry observed here. Unfortunately, there are
no spatially resolved observations of the molecular emission of this disk that would attest to the similarity of this
structure with GoHam b. 

Near-IR observations of scattered light from edge-on disks are generally characterized by asymmetric 
disk structures that are often attributed to illumination effects. The most striking example of such an asymmetry 
is observed in the HH 30 disk \citep{sta99}. The fluorescence of UV-excited
PAHs at IR\ wavelengths dominates the emission, hence UV illumination effects could also create the observed 
asymmetry. While we cannot exclude this possibility, the scenario of illumination effects cannot
explain the coincidence of the mid-IR emission decrease with the position of the molecular emission.
{We also note that illumination effects are usually variable in time in such systems, and the Hubble space telescope 
images of GoHam obtained in the visible and near-IR do not show evidence of this variability
(see images in \citealt{buj08} and \citealt{woo08}).}

%The large scale-height observed in PAH emission places GoHam
%in the category of a transition disk \citep{mee01} which are now all thought to be transition disks 
%with large gaps ($\sim$ 50 AU) created by the presence of a massive planet, which gives rise to a 
%huge wall at a large distance, resolved at e.g., mid-IR wavelengths (cf., Maaskant et al and others). 
%GoHam could be an extreme example of such a system with a gap of $\sim$ 350 AU. Given the observed 
%asymmetry in gas and PAHs, and the derived large mass associated with this asymmetry (Sect.~\ref{sec_prop}),
%it is tempting to speculate that this is the location of a giant planet in nascence.

\subsection{Comparison to candidate protoplanets}

Recently, \citet{qua13} identified a point source in the disk around HD 100546, a Herbig star of B9 spectral type.
These authors suggested that this source might represent early stages of planetary formation
and called the object HD 100546 b. This source shares similarities with GoHam b since
it appears as an isolated clump inside a massive disk around an intermediate-mass star. 
The mass of HD 100546 b is only poorly constrained, but is estimated to range between 
1 and 15 M$_{Jup}$ \citep{qua13, boc13, cur14}, which is similar to GoHam b. 
However, GoHam b is situated at a much larger distance ($330\pm70$ AU) from its host
star than HD 100546 b ($\sim$ 50 AU). In addition, \citet{qua14} derived an effective 
temperature of $\sim$ 1000 K for HD 100546 b, while GoHam b appears to be a rather
cold object (at most a few tens of K according to the CO emission). While the orbit of 
GoHam b may seem large,  it is not exceptional in the context of the recent discovery of 
a 5 M$_{Jup}$ planet separated from its host star HD 106906 by 650 AU \citep{bai14}. Finally, it should be noted that 
\citet{dut14} also reported a clump of molecular gas at the outer edge of the disk ring around GG Tau A, 
at a radius of $\sim$ 250 AU. Although this latter clump is poorly studied, it appears to be quite similar
to GoHam b.

\subsection{A young protoplanet resulting from a GI ?}

Since both HD 100546 b and GoHam b are found at large radii and inside massive disks, a natural 
interpretation for their origin is the gravitational instability. 
%However, the latter authors noted that the disk around HD 100546 does not meet the theoretical 
%criteria for GI to occur. We have evaluated the stability of GoHam's 
%disk again gravitational instability using the classical Toomre $Q$ parameter \citep{too64}. For the physical
%conditions as described in Table \ref{table1}, we find that $Q>1$ at all radii, i.e. that the disk is stable.
%However, since the density (a critical parameter in the disk stability) in Table 1 was derived using 
%CO observations, and since CO can be largely depleted onto dust grains in the disk mid plane, 
%it may be underestimated by a large factor and the stability overestimated. 
%A density estimate
%based on the dust optical depth in the mid plane indeed yields a minimal density of 10$^{7}$ cm$^{-3}$
%In addition, if GoHam b is the results of a GI, the disk could have been unstable in the past 
%while it is now stable.
Hydrodynamical models indicate that GI clumps have a typical size of $\sim 0.4\times H$ 
\citep{bol10}, where $H$ is the disk scale height inferred from the midplane temperature and the 
angular velocity. 
Assuming a minimal orbit of $350\pm50$ AU for the clump and using the disk parameters in 
Table \ref{table1} yields a { minimum radius of $r=0.4H=30\pm5$ AU}, on the same order of magnitude as the value 
derived from the model: $r_b=135\pm15$ AU (see Appendix~\ref{app_mod}).
Using Eq. (12) of \citet{bol10}, we  derive a theoretical mass for a GI clump at 
$350\pm50$ AU of 5.5$\pm1$ M$_{Jup}$, which falls in the mass range derived 
in Sect.~\ref{sec_prop} of $0.8-11.4$ M$_{Jup}$. 
We can estimate the midplane density $n_I$ required for the disk to be unstable to the GI 
at the radius corresponding to the angular separation at which GoHam b is observed, 
using the classical Toomre $Q$ parameter \citep{too64}. For the physical conditions in Table \ref{table1}, 
we find that the disk is unstable, that is, that $Q\lesssim 2$ for a density higher than 
$n_I\sim 2\times10^8$cm$^{-3}$, in agreement with the observed lower limit $n>10^6$cm$^{-3}$ 
derived from CO observations (Table \ref{table1}).
%Theoretical models predict that the ratio between the clump and background density
%is of the order of 10 \citep{paa12}, implying a minimum density in the disk midplane larger than 
%0.1$\times n_b$, i.e. $3.7\pm0.7\times10^6$ cm$^{-3}$, also in consistent with this observational lower limit.

\section{Conclusion}

Overall, GoHam b consists of a small (of about 100 AU) and dense (of about $10^{7}$ cm$^{-3}$) 
structure of molecular gas and dust, with a mass of $0.8-11.4$ M$_{Jup}$ (for a dust-to-gas mass ratio of 0.01).
These results are fully consistent with those of \citet{buj09}, which in addition have shown that the position of
GoHam b corresponds to a modification of the Keplerian velocity field. 
Altogether, this evidence indicates that GoHam b is a promising candidate of a protoplanet 
formed by gravitational instability. Additional studies, in particular at higher angular resolution and with other tracers using 
ALMA coupled to detailed 3D radiative transfer modeling, are needed to confirm this hypothesis and 
to rule out other possibilities (such as the presence of an arc of gas and dust).

\begin{acknowledgements}
This work was supported by the CNRS program ``Physique et Chimie du Milieu Interstellaire" (PCMI). 
P. Pilleri acknowledges financial support from the Centre National d'Etudes Spatiales (CNES). 
\end{acknowledgements}
%\begin{thebibliography}{}

\bibliographystyle{aa}
\bibliography{biblio}

 %\end{thebibliography}
%{\appendix \section{Temperature of the atomic gas} \label{appendix}
%It is possible to estimate a lower limit for the temperature of the warm gas component
%traced by PAHs, assuming vertical hydrostatic equilibrium. In this situation,
%the warm disk scale height $H_{PAH}$ can be related to the sound speed 
%--and hence temperature-- of the gas through 
%$H_{PAH}=\sqrt{2}~c~\Omega_K^{-1}$,
%where  $c$ is the sound speed given by  $c=\sqrt{5/3~k~T_{gas}^{PAH}/\mu m_H}$ 
%with $m_H$ the hydrogen atom mass, $\mu$ the reduced mass coefficient, and
%$T_{gas}^{PAH}$ is the temperature of the gas where PAHs are emitting.
%$\Omega_K=\sqrt{G.M_*./R_{PAH}^3}$ is the Keplerian velocity. 
%Using the parameters in Table 1 yields $T_{gas}^{PAH} >150$~K,
%which is in agreement with models of highly irradiated PDRs 
%which predict temperatures of a few 100 K in the atomic gas.

\newpage

{\appendix
\section{Supplementary figure}
\begin{figure}[ht!]
\begin{center}
\includegraphics[width=9cm]{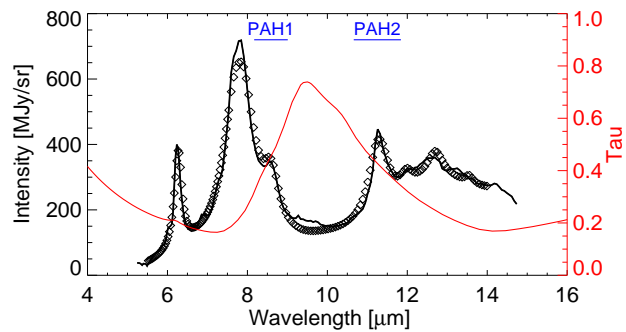}
\vspace{-0.0cm}
\caption{ Spitzer IRS spectrum of GoHam (continuous black line) and fit (diamonds) using the PAHTAT model. The mid-IR optical depth derived with the model is shown in red. The positions and bandwidths of the VISIR PAH1 and PAH2 filters are indicated. This spectrum includes GoHam a and b, the absorption is dominated by the disk mid-plane.
 \label{fig_spitzer}}
\end{center}
\vspace{-0.5cm}
\end{figure}

\section{Model}\label{app_mod}

\subsection{Model description}
We modeled GoHam b as a spherical clump of constant 
density $n_b$ and radius $r_b$. 
The presence of this clump results in a localized excess of column density 
with a profile along the cut as a function of position $p$ (Fig.~\ref{fig_cut}) 
of parabolic form:
\begin{equation}
N_{H}(p)={n_b}\times 2~\Re \left( \sqrt{r_b^2-(p_b-p)^2} \right)
,\end{equation}
where $p_b$ is the position of GoHam b
in the cross cut, that is, 4.25''.
The resulting mid-IR optical depth $\tau_b$ due to GoHam b is then
\begin{equation}
\tau_b(p)=C_{ext}\times N_H(p),
\end{equation}
where $C_{ext}$ is the dust cross-section per unit of H atom.
The mid-IR emission along the cut, corrected for the extinction
by GoHam b, can then be recovered by \begin{equation}
I^0_{IR}(p)=I_{IR}(p) /\exp{(-\tau_b(p))},
\label{eq_ir}
\end{equation}
where $I_{IR}(p)$ is the observed mid-IR emission profile shown in Fig.~\ref{fig_cut}.
%$I_0$ can be estimated easily
%by the following approach: since the Northern part of the disk
%is not affected by GoHam b ()
In addition, we computed the emission in the $^{13}$CO(2-1)
line as a function of $p$:
\begin{equation}
{I_{CO}(p)}=T_{CO}^{peak}(p) \Delta v~\Omega_{ff}\otimes K,
\end{equation}
where $ \Delta v$ is the width of the line measured to be 1.7 km/s, $\Omega_{ff}$ is the beam-filling factor
equal to $ (a\times b) /r_b^2$, where $a$ and $b$ are the minor and major axis of the beam, that is,
$a=1.14''$ and $b=1.52''$.  $K$ is a Gaussian kernel of width equal to $b$, which is the angular resolution
along the cut shown in Fig.~\ref{fig_cut} (since the major axis of the beam is almost aligned north-south,
see \citealt{buj09}, as is our cross cut). $T_{CO}^{peak}(p)$ is the brightness
temperature of the  $^{13}$CO(2-1) line that is equal to zero for $p<p_b-r_b$ and $p>p_b+r_b$.
For the other values of $p$, the brightness is equal to (optically thick line)
\begin{equation}
T_{CO}^{peak}=\frac{h\nu}{k}\times\frac{1}{(\exp(h\nu/(kT_{ex}))-1)},
\end{equation}
where $T_{ex}$ is the excitation temperature, which we fixed at 16 K,
that is, the temperature of the disk midplane derived by \citet{buj09}.
$T_{ex}$ could be higher than this value in the internal parts of the clump,
but this is not critical in the estimation of $r_b$ , which is a function of the square root
of $T_{ex}$. Overall, the parameters of the model are the density of GoHam b
$n_b$, the radius of GoHam b $r_b$, and
the dust cross-section $C_{ext}$. For interstellar dust, the cross section is typically 
$C_{ext}=2.5\times10^{-23}$ cm$^2$ per H atom at 8.6 and 11.2 $\mu$m \citep{wei01}. 
In disks, however, this value is expected to decrease significantly due to grain growth, 
typically by a factor of 10 in the mid-IR \citep{and09, dal01}. We therefore considered these
two extreme cases and their effects on the parameters derived by the model.

\subsection{Adjustment of the parameters $n_b$ and $r_b$}

First, $r_b$ was adjusted so as to reproduce the observed emission profile 
of $I_{CO}$ as shown in Fig.~\ref{fig_cut}. Once the value
of $r_b$ is adjusted, the following step consists of adjusting  the parameter $n_b$
so as to obtain a symmetric $I^0_{IR}(p)$ profile, which is what is 
expected in a disk without any clump. The result of this procedure is
shown in Fig.~\ref{fig_cut}. 
%\subsection{Model outputs}
From the adjusted values of $n_b$ and $r_b$, we can derive the clump
mass,
\begin{equation}
M_b=4/3~\pi~r_b^3 \times n_b~\mu~m_H,
\end{equation}
where $\mu$ is the mean molecular weight and $m_H$ the proton mass.
The parameters used in the model and those derived from the fit for the two
values of $C_{ext}$ are summarized in Table~\ref{tab_mod}.

%\newpage
%\section{Supplementary Tables}

 \begin{table*}[ht!]
\caption{Main physical parameters of the GoHam disk. Error bars result from the uncertainty on the distance to GoHam.}
\label{table1}
\begin{center}
\begin{tabular}{lccc}
\hline
Parameter                       &                               & Comment                         & Ref.\\
\hline
\hline
\multicolumn{4}{c}{GoHam a }\\
\hline
$d$                                     & $250\pm50$ pc                         &         Distance                                & (1,2, 3) \\
$M_*$                           & $2.5 \pm 0.5$ $M_{\sun}$      &          Mass of the star$^1$   & (1) \\
$R_{PAH}$                       & $750\pm150$ AU                        &         See Fig. 2                              & (4) \\ 
$R_{mol}$                       & $1650\pm350$ AU                       &         See Fig. 2                              &(1) \\
$H_{PAH}$                       & $775\pm 150$ AU                       &         See Fig. 2                              &(4)\\
$H_{mol}$                               & $450\pm90$ AU                 &         See Fig. 2                              &(1) \\
$n$                                     & $> 10^6$ cm$^{-3}$            &         Midplane density$^2$    &(1) \\
$T$                                     & 16 K                                  &         Midplane gas temp.$^3$  &(1) \\
\hline
\multicolumn{4}{l}{\tiny(1) \citealt{buj09} (2) \citealt{ber09} (3)\citealt{woo08} (4) This work }\\
\multicolumn{4}{l}{\tiny $^1$ From the Keplerian velocity field. $^2$ Lower limit from CO observations}\\
\multicolumn{4}{l}{\tiny $^3$ Assumed to be uniform with radius. }\\
\end{tabular}
\end{center}
\end{table*}

 \begin{table*}[ht!]
 \label{tab_mod}
\caption{Main physical parameters of the GoHam b candidate protoplanet in the model. Error bars result from the uncertainty in adjusting the model  and on the distance to GoHam (the latter being dominant). The values in parenthesis
correspond to the case when $C_{ext}$ is a factor of 10 smaller than the ISM value, i.e., $C_{ext}= 2.5\times10^{-24}$ cm$^{2}$/H. }
\begin{center}
\begin{tabular}{lll}
\hline
%Parameter                                      & Comment\\
\hline
\multicolumn{3}{c}{Input parameters}\\
\hline
$C_{ext}$                               & $2.5\times10^{-23}$                 ( $2.5\times10^{-24}$)          & Mid-IR dust cross-section in cm$^{2}$/H \\
$p_b$                           &       4.25                    & Position of GoHam b in cross cut Fig.~\ref{fig_cut}     ('')                             \\
$\Delta v$                              &       1.7                             & Width of the $^{13}$CO line in km/s                             \\
$T_{ex}$                                &       16                              & CO excitation temperature in K                  \\
\hline
\multicolumn{3}{c}{Output parameters}\\
\hline
$r_b$                           &  $155 \pm31   $               & Clump radius     in AU  \\
%$N_H^{peak}$           & 4.5$\times10^{22}$    (4.5$\times10^{23}$) & Peak column density in        cm$^{-2}$               \\
$M_b$                           & $0.95\pm0.19$         (9.5$\pm$1.9)                 & Clump mass in Jupiter masses $^1$          \\
$n_b$                           & $7.0\pm1.4\times10^6$         ($7.0\pm1.4\times10^7$) & Clump density in cm$^{-3}$              \\
\hline
\multicolumn{3}{l}{\tiny $^1$ For a dust-to-gas mass ratio of 0.01. }\\
\end{tabular}
\end{center}
\end{table*}
}

%\section{Density in the disk midplane}

%We can derive the minimal density in the mid plane using the following approach.
%The PAH1 emission peaks at a value of the order of 0.15 Jy/arcsec$^{2}$. In the disk
%mid plane, the PAH1 emission drops down to a value of about 0.02 due to reabsorption
%by the disk. This implies an optical depth at 8.6 $\mu$m of $\tau$=2. Therefore,
%following the approach described in Sect.~\ref{}, this yields a column density of 
%$2.5\times10^23$ cm$^{-2}$, which divided by the disk radius of 2000 AU (table 1)
%yields an hydrogen density of $10^7$ cm$^{-3}$. }

\end{document}